Kuen Yao Lau[1,*]

# Recent advances of MXene saturable absorber for near-infrared mode-locked fiber laser

[1]State Key Laboratory of Modern Optical Instrumentation, College of Optical Science and Engineering, Zhejiang University, Hangzhou 310027, China.

*0621072@zju.edu.cn

**Abstract:** To date, MXene has been discovered for its viability as alternatives to conventional saturable absorber such as carbon nanotube and graphene. The characteristics of high nonlinear saturable absorption, astounding modulation depth, flexible bandgap tunability, and high electron density near Fermi level are the fundamentals of the MXene as an excellent saturable absorber candidate. In particular, the research effort contributed to MXene in nonlinear ultrafast optics are extensively growing because MXene comprises one of the largest families in 2D nanomaterials that provides huge combination possibilities by forming a class of metal carbide or metal nitride with 2D layered structure. Herein, this review summarizes the recent development on synthesis and material characterization of the MXene, the studies on its nonlinear saturable absorption and the application of the MXene saturable absorber in near-infrared mode-locked fiber laser. Finally, some issues and challenges as well as future perspectives of this novel material are discussed.

## 1   Introduction

Nonlinear optics describes the interaction of light with matter. In particular, nonlinearity gives the insight for the response of the materials with respect to the amplitude of the applied electromagnetic field [1]. The nonlinearity effect experiences a rapid proliferation of interesting optical phenomenon, such as saturable absorption that could generate ultrashort pulsed laser in picosecond or femtosecond regime. A saturable absorber (SA) exhibits intensity-dependent transmission, whereas the optical loss reduced with higher light intensity [2]. The SA is a passive optical device to generate ultrashort pulsed laser via passively mode-locking mechanism in a fiber laser cavity. The development of fiber laser with ultrafast characteristics has gained marked interest due to the advantages of system compactness, cost-efficient, high beam quality, and excellent heat dissipation of the optical fiber [3, 4]. Subsequently, ultrafast fiber laser with pulse duration faster than picosecond shows potential in applications such as ophthalmology [5], bio-imaging [6], advanced micro-machining [7], quantum information processing [8], molecular spectroscopy [9], and nonlinear microscopy [10]. In near-infrared (NIR) region, there are several wavelengths feasible for practical applications, such as high-power ultrafast laser at 1064 nm [11], telecommunication with low attenuation and long-haul application at 1550 nm [12], and tissue cutting for medical application due to high water absorption at ~1950 nm [13]. In addition, ultrafast laser in ~1950 nm can suppress coagulation of blood in surgery applications, as well as for mechanical processing such as cutting, welding and marking due to the high absorption of typical plastic at this wavelength region [13].

There are numerous SAs reported for the generation of ultrashort pulsed laser since the first demonstration of femtosecond pulse with intracavity saturable absorber dye cell, which is a flashlamp-pumped dye laser in 1974. A semiconductor saturable absorber mirror (SESAM) is the most commonly used SA technology that have become prevalent materials for the research and market for a long time. The SESAM composes of numerous layers of semiconductor materials grown via atomic layer deposition or other microfabrication technique which typically has a reflective mirror within these material layers for the free-space alignment of a fibre laser cavity [14]. Therefore, the fabrication challenges and the difficulties for the integration of an all-fibre laser cavity are two major drawbacks for the SESAM. In addition, poor beam quality, low damage threshold, and narrow operating bandwidth within 10 nm limits its wide applications [15]. After that, carbon nanotube (CNT) SA was firstly proposed by Set et al. in 2004 for passively mode-locked laser. The CNT addresses most of the limitations in SESAM with numerous advantages such as high nonlinear optics (NLO) susceptibilities, low saturation intensity, fast recovery time and sub-picosecond relaxation time [16-18]. However, the difficulties in chirality and tube diameter control of CNT restricts its integration in longer working wavelength such as 2 μm [19, 20]. After that, graphene SA was firstly demonstrated after CNT in 2009 by Hasan et al. [21] and Bao et al. [22]. Graphene is a 2D band-free semi-metal with semiconducting properties and exhibits ultrafast re-covery time, electron mobility up to 106 cm$^2$V$^{-1}$S$^{-1}$, and strong nonlinear refractive index of ~10$^{-7}$ cm$^2$/W [23-25]. In spite of graphene has inherent zero bandgap for broadband operation ranging from X-ray, ultraviolet, visible, to near- and mid-infrared, its low on/off switching ratio and low absorption of 2.3% per graphene layer restricts its laser output performance and practical applications for optoelectronics [26, 27]. Transition metal dichalcogenides (TMD) were demonstrated after graphene with the advantages of scalable bandgap properties via flexible defect engineering [28]. For instance, TMD has both semiconductor and superconducting properties, tunability from indirect bandgap to direct bandgap, wide range of electronic band structures, nonlinear absorption characteristics, as well as large modulation depth by reducing its thickness [29-32]. Nevertheless, direct bandgap characteristic only occurs typically in monolayer TMD which requires a complicated synthesis and defect engineering process [33, 34]. Recently, the exfoliation of mono-elemental group-V materials such as phosphorus and bismuth from its bulk crystal into phosphorene and bismuthene has been demonstrated as suitable candidates of the SA to generate ultrashort pulse in near-IR region [35-37]. These mono-elemental materials tune the bandgap and properties through transition between indirect to direct bandgap by changing the material layers [38, 39]. However, the instability to oxidation remains a hurdle to make them an environmentally stable SA [40, 41]. Between these two materials, bismuthene has better resistivity to oxidation than phosphorene [42].

Apart from the aforementioned SA materials, the MXene was firstly discovered for the demonstration of 2D Ti$_3$C$_2$ nanosheets, multi-layer structures and conical scrolls by Y. Gogotsi in 2011 [43]. Since then, this material has received a rapid proliferation of research interest in many areas such as actuator [44], sensor [45], clean water production [46], photothermal therapy [47], and lithium-sulphur batteries [48, 49]. MXene comprises one of the largest families in 2D nanomaterials that form a class of metal carbide or metal nitride with 2D layered structured that the materials are stacked on top of each other [50]. The general formula for MXene materials is $M_{n+1}X_nT_x$, where n=1,2,3. M is the transition metals (Sc, Ti, Zr, Hf, V, Nb, Ta, Cr, Mo), X is carbide, nitride, or carbonitride, T is an element from the 3$^{rd}$ or 4$^{th}$ main group (-OH, -O, or -F), and x denotes the number of terminal groups [51, 52]. In the typical structure of MXene, n+1 layers of M covers n layers of X in the form of [MX]$_n$M. According to the variation of chemical elements, atomic structures and terminated group species, MXene can takes the form in hundreds of 2D material members. In contrast to other 2D materials where the stacking is occurred through Van der Waals' force, the stacking of MXene is primarily formed by the hydrogen bonds between surface functional group or indirect hydrogen bonds together with intercalated water molecules [53, 54]. This contributes to higher stacking energy of MXene such as Ti$_3$C$_2$T$_x$ at 56.691 meV/Å$^2$ which is 8.2 times magnitude larger than graphene at 6.898 meV/Å$^2$ [13]. Additionally, MXene has lower linear absorption than graphene [55], large contacting surface area with hydrophilic characteristic [56, 57], high modulus of elasticity [58], excellent electrical conductivity and stability [59-61], high electron density near Fermi level [62], as well as tunable optoelectronics properties [63, 64]. Aside from these advantages, MXene possesses excellent nonlinear saturable absorption such as Ti$_3$C$_2$T$_x$ has nonlinear absorption in two order of magnitude higher than molybdenum disulfide (MoS$_2$) and black phosphorus (BP) that the transmission is increased with higher light fluences [65], astounding modulation depth as high as ~58.41% [66], excellent mechanical performance [67], large bandgap tunability [68, 69], excellent thermo-electric properties [70], and environmental stable at humidity condition [71]. Moreover, monolayer MXene with thickness of ~1 nm and lateral sizes in the order of µm is also feasible by delamination [52].

A recent review for MXene and other 2D materials was done in terms of synthesis, optical properties and applications in ultrafast photonics by Zhang et al. [72]. Although MXene appears to be a relatively new material compared to graphene, TMD, black phosphorus and other materials, it has started to attract the research interest as an excellent SA candidate. Herein, this review will focus on the latest development of MXene SA to generate mode-locked fiber laser in near-infrared wavelength region. The studies are expected to bring better insight for the readers on the exploration of MXene-based saturable absorber in passively mode-locked fiber laser, that could also generate ideas for the researches in conjunction with the numerous demonstrations, issues and challenges as well as the future perspectives that were brought upon by these reports.

## 2 Synthesis and characterization of MXene

MXene comprises of various chemical forms depending on the M, X and T elements in the periodic table. To date, most MXene synthesis was conducted through top-down approach, particularly on the selective etching of their MAX phases [73]. In this section, a review was done on the synthesis, material properties and the characterization for MXenes which includes Ti$_3$C$_2$T$_x$, Ti$_3$CNT$_x$, V$_2$CT$_x$, and Nb$_2$C as the saturable absorber.

## 2.1 Ti$_3$C$_2$T$_x$

The most popular MXene Ti$_3$C$_2$T$_x$ has been predominantly deployed, overwhelming other MXenes in optoelectronics applications due to its narrow bandgap and high photothermal effect [67, 74-76]. The Ti$_3$C$_2$T$_x$ is constructed from five atomic layers Ti-C-Ti-C-Ti, whereas the top and bottom Ti layers are significant for the surface termination [13]. Meanwhile, the intrinsic existing surface termination plays the role as a buffer-like separator between the Ti$_3$C$_2$T$_x$ profile. Naguib et al. [43] has theoretically predicted the bandgap of Ti$_3$C$_2$T$_x$ could be less than 0.2 eV, which is thus feasible for broadband response. For instance, the NLO response of Ti$_3$C$_2$T$_x$ was reported with the range from 800 nm to 1800 nm [77-79], and also lately reported with larger range from 400 to 2200 nm [13] and 200 to 1950 nm [55]. Additionally, the NLO absorption of the Ti$_3$C$_2$T$_x$ was reported with range from 1.9 to 2 µm [13, 80], mid-infrared of ~2.8 µm [62, 81, 82], as well as microwave and terahertz [83, 84]. Apart from that, the Ti$_3$C$_2$T$_x$ has high nonlinear absorption coefficient of ~$10^{-13}$ e.s.u. and large effective nonlinear absorption coefficient ($\beta_{eff}$) of ~-$10^{-21}$ m$^2$/v$^2$ [77]. This denotes its strong optical switching capability, graphene alike layer structure with interlayer distance of 0.98 nm, and the $\beta_{eff}$ is one order of magnitude larger than graphene thus interpreting its huge potential as SA, modulator, and optical switches. The Ti$_3$C$_2$T$_x$ was experimentally presented with excellent saturable absorption characteristics such as modulation depth of ~50% and more resilient than other 2D materials with higher optical damage threshold of ~70 mJ/cm$^2$ [65]. However, the Ti$_3$C$_2$T$_x$ has several limitations. For instance, the surface termination preserves the characteristics of monolayer MXene to be influenced by adjacent monolayer thus additional engineering procedures are needed to regulate the its electronic and optical properties [43, 74, 85]. Moreover, the open vials Ti$_3$C$_2$T$_x$ solution degrade into cloudy-white colloidal solutions containing mainly of anatase (TiO$_2$) which is impractical for many applications [86, 87].

The synthesis of Ti$_3$C$_2$T$_x$ was prepared based on the basis of the aqueous acid etching method reported by Naguib et al. [43]. Based on Naguib's method, the Ti$_3$AlC$_2$ was synthesized as the final product instead of Ti$_3$C$_2$T$_x$. Nonetheless, the aqueous acid etching procedures are useful references to synthesize Ti$_3$C$_2$T$_x$. In Naguib's work [43], the raw materials Ti$_2$AlC and TiC were ball-milled and the mixture was heated to 1350 °C for 2 hours under argon gas condition. Next, ~10g of the powder was immersed with 40 wt.% hydrofluoric (HF) acid in a volume ratio of 1:10 at room temperature for 2 hours. The attained deposit was flushed with DI water and centrifuged to separate the powders. Similarly, the synthesis of Ti$_3$AlC$_2$ sample was also presented through aqueous acid etching method in [52]. Subsequently, the Ti$_3$AlC$_2$ powder was mixed with 40 wt.% HF acid for 48 hours at room temperature, washed with DI water for several times and the resulting deposition was dried in vacuum oven at 60 °C for 48 hours. Besides that, a Ti$_2$AlC MXene was recently demonstrated for its better performance than Ti$_3$AlC$_2$ such as higher stability [88], lower Fermi level to achieve electronic band transition [89], higher optical conductivity [90], and better resistance to oxidation [91]. Nonetheless, this Ti$_2$AlC was directly obtained from the commercially available product without going through the synthesis process [89]. The Ti$_2$AlC was made into a composite with polyvinyl alcohol (PVA) in a ratio of 1:1. The characterization of this Ti$_2$AlC was very thoroughly investigated such as absorption spectrum, Raman spectrum, SEM image, EDS and XRD spectrum.

Aside from Ti-Al-C MXene, there are several works demonstrated for the preparation of Ti$_3$C$_2$T$_x$ dispersion via the aqueous acid etching method. The preparation of Ti$_3$C$_2$T$_x$ flakes was done through immersion of the MAX phase Ti$_3$AlC$_2$ with 40% HF that was stirred with 500 rpm for 20 hours at room temperature [82]. Next, the bulk Ti$_3$C$_2$T$_x$ solution was obtained and washed with DI water until the pH is more than 6. Finally, the Ti$_3$C$_2$T$_x$ flakes were obtained after probe-ultrasonication with cell grinder for 24 hours. The preparation of Ti$_3$C$_2$T$_x$ suspension was also conducted with HF etching by using isopropanol as the solvent, and its thin film was prepared through liquid phase exfoliation (LPE) method [78]. Fig. 1(a) shows the Raman spectrum of the prepared Ti$_3$C$_2$T$_x$ suspension characterized with a 532 nm laser. Based on the characteristic peaks, the 267 cm$^{-1}$ is contributed by the H atoms in the OH groups of Ti$_3$C$_2$T$_x$, the 565 cm$^{-1}$ peak is due to the O atoms A$_{1g}$ vibration, and the 797 cm$^{-1}$ is attributed to the Ti-C and C-C vibrations [78]. In addition, the Raman peaks at 1093, 1346 and 1589 cm$^{-1}$ are the characteristics peaks of the mixture between ethanol and isopropanol applied in the synthesis process [92]. In [63], aqueous acid etching method was employed by mixing Ti$_3$AlC$_2$ solution with 40% HF using a volume ratio of 1:15. The obtained deposit was flushed with DI water to attain pH of greater than 6 which was then dried in a vacuum oven. Next, water bath sonication was employed to exfoliate the Ti$_3$C$_2$T$_x$ by dispersing the Ti$_3$C$_2$T$_x$ powder in IPA at a temperature below 20 °C for 10 hours. Next, the suspension liquid was centrifuged at 3000 rpm for 30 minutes and the supernatant was then centrifuged at 18000 rpm for 30 minutes. After removing the supernatant, the precipitate was dispersed in DI water to obtain the Ti$_3$C$_2$T$_x$ suspension. Fig. 1(b) shows the SEM image of the delaminated Ti$_3$C$_2$T$_x$ after HF etching. The accordion-like structure indicates excellent stripping effect after the etching process. Fig. 1(c) illustrates the TEM image of the Ti$_3$C$_2$T$_x$ which reveals the measured nanosheet size ranging from 50 to 200 nm. Based on the AFM measurement, the thickness of the nanosheet is ~2 to 3 nm for the six selected dashed lines as illustrated in Fig. 1(d) to Fig. 1(f).

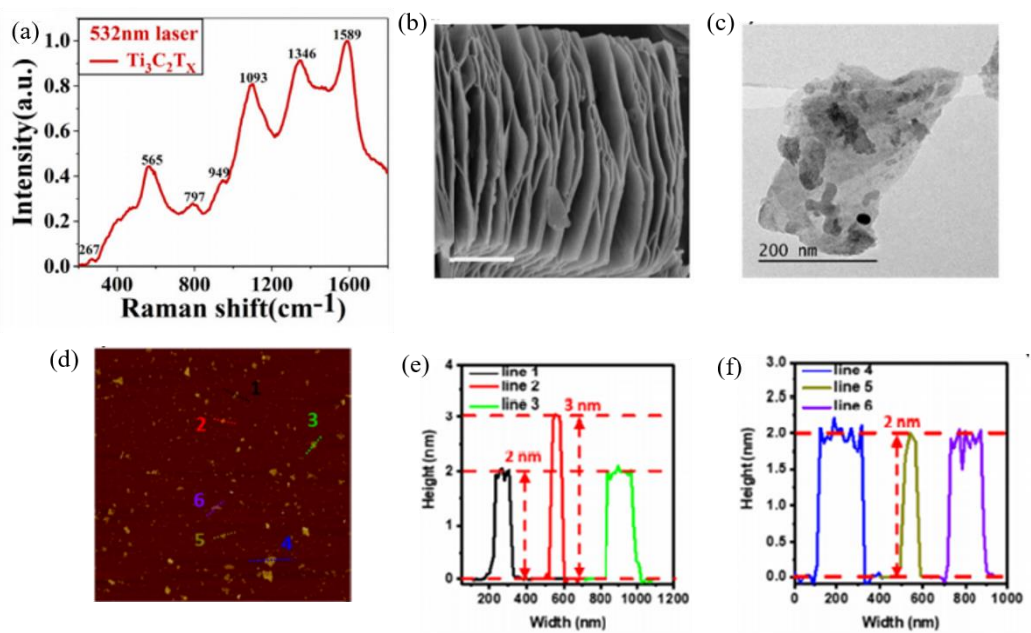

**Fig. 1:** (a) Raman spectrum of $Ti_3C_2T_x$ suspension, J. Feng et al. [78]. © WILEY-VCH Verlag GmbH & Co. KGaA, Weinheim 2020. (b) SEM image, (c) TEM image, and (d)-(f) AFM measurement for $Ti_3C_2T_x$ prepared with aqueous acid etching process, Q. Wu et al. [63]. © Optical Society of America 2019.

Another aqueous acid etching process for the synthesis of $Ti_3C_2T_x$ nanosheet solution was demonstrated in [77]. The deposit was firstly dried in the vacuum furnace and the $Ti_3C_2T_x$ powder was dissolved in 1 mg/mL of N-methyl-2-pyrrolidone (NMP). Next, the supernatant was centrifuged with a speed of 4000 rpm for 20 minutes at 10 °C to attain the $Ti_3C_2T_x$ nanosheet solution. Similar synthesis method was also presented in [81]. The SEM image of the prepared $Ti_3C_2T_x$ powder is depicted in Fig. 2(a). The accordion-like structure denotes the good extraction of Al atoms from the corresponding MAX phase [93]. The interlayer thickness of this $Ti_3C_2T_x$ powder was observed via the HRTEM image in Fig. 2(b). From the image, the interlayer distance was measured to be ~1 nm. The atomic lattice was characterized with hexagonal space group $P6_3/mmc$ as shown in Fig. 2(c). The mean particle size of the $Ti_3C_2T_x$ powder was measured as 21.5 ± 7.6 nm, whereas the thickness of the $Ti_3C_2T_x$ powder was deduced as 3.7 ± 0.7 nm from the AFM measurement as depicted in Fig. 2(d). The zero bandgap of $Ti_3C_2$ ($E_g < 0.2$ eV for $Ti_3C_2T_x$) proves its viability for broadband operation [51, 94]. In addition, aqueous acid etching method shows advantages in the aspects of feasibility, yielding, controllability, and cost efficiency than bottom-up CVD method [81].

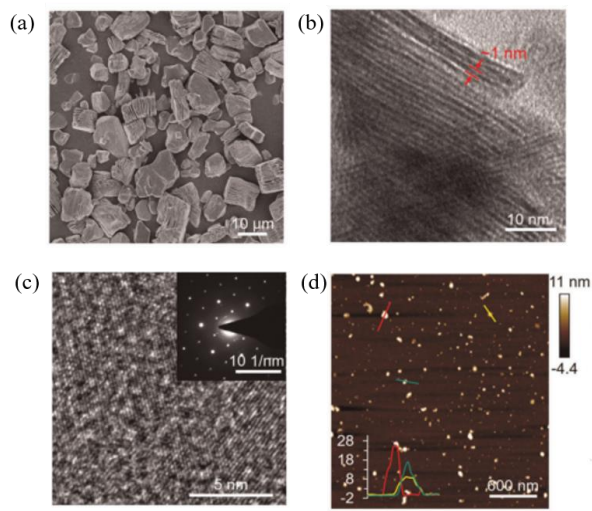

**Fig. 2:** (a) SEM image of the prepared MXene powder via aqueous acid etching method, (b) HRTEM image shows the interlayer thickness of MXene, (c) HRTEM image of the MXene nanosheet, and (d) AFM image of the MXene nanosheet in IPA ink precursor, X. Jiang et al. [81]. © Springer Nature 2019.

The termination of the functional groups on the MXene surface is expected to offer better stability and superior properties to the MXene [95, 96]. Nonetheless, the integration of HF in the aqueous acid etching process synthesizes MXene with contamination by oxygenic and hydrophilic termination groups, such as -O and -OH on the MXene surface which degrades both the environmental stability and material properties of the MXene [64, 86]. The layered MAX phases are typically exfoliated via aqueous HF etching, whilst HF is a highly-acidic solvent which causes the formation of defects on the MXene surface as well as severe pollution [97-99]. This safety issue could be overcome by a rather simpler and safer synthesis process based on electrochemical exfoliation that implement highly fluorinated, non-aqueous ionic liquid as the electrolyte to prepare $Ti_3C_2T_x$ quantum dot (QD) with enhanced stability [64]. The synthesis tools are constructed from a three-electrode electrometrical system consisting of a working electrode (bulk $Ti_3AlC_2$), a quasi-reference electrode (Ag wire) and a counter-electrode (Pt wire) as shown in Fig. 3(a). The non-aqueous electrolyte was prepared by mixing 20 g of [EMIM][$PF_6$] into 100 mL of MeCN. Next, the mixed solution was poured inside a 150 mL container consisting of three aforementioned electrodes and a glass pipe. Afterwards, the mixed solution electrolyte was bubbled with argon gas for 5 minutes before use. A constant potential of 3-7 V was applied to the Ag wire and the process was performed for 5 hours. Then, the electrolytic decomposition was deployed to decompose $PF_6^-$ into $F^-$, meanwhile the Al layer was selectively etched away from the $Ti_3AlC_2$. After the etching process, the electrolyte appears brownish and was suspended with $Ti_3C_2T_x$ powder and fragments. Next, two consecutive centrifugation processes were conducted, the first 30 minutes at 3500 rpm for the separation of large particles from the electrolyte and the second 30 minutes at 7000 rpm to obtain the sediment. After that, the sediment was added into the MeCN and ultra-sonicated under $N_2$ gas ambient condition for 10 hours. Finally, the $Ti_3C_2T_x$ QD was obtained by removing the impurities of some large particles from the supernatant through further centrifugation at 7000 rpm for 30 minutes. Fig. 3(b) illustrates the TEM image of the synthesized $Ti_3C_2T_x$ QD with uniform and ultra-small properties. Subsequently, the $Ti_3C_2T_x$ QD was examined with its crystal structure via the HRTEM as shown in Fig. 3(c) and 3(d). According to Fig. 3(d), the lattice fringes have an inner plane spacing of 0.21 nm in conjunction with the (0110) facet of the $Ti_3C_2$. In addition, the average lateral size of the $Ti_3C_2T_x$ QD was deduced as 5.34 nm according to the statistical TEM analysis of 100 $Ti_3C_2T_x$ QDs as depicted in Fig. 3(e). Based on the XRD pattern of the $Ti_3C_2T_x$ QD in Fig. 3(f), the diffraction peaks with 2θ values at 9.5°, 18°, 28°, 39°, and 31.8° correspond to the (002), (004), (008), (101), and (105) crystal planes of crystalline $Ti_3AlC_2$, respectively [75, 100, 101].

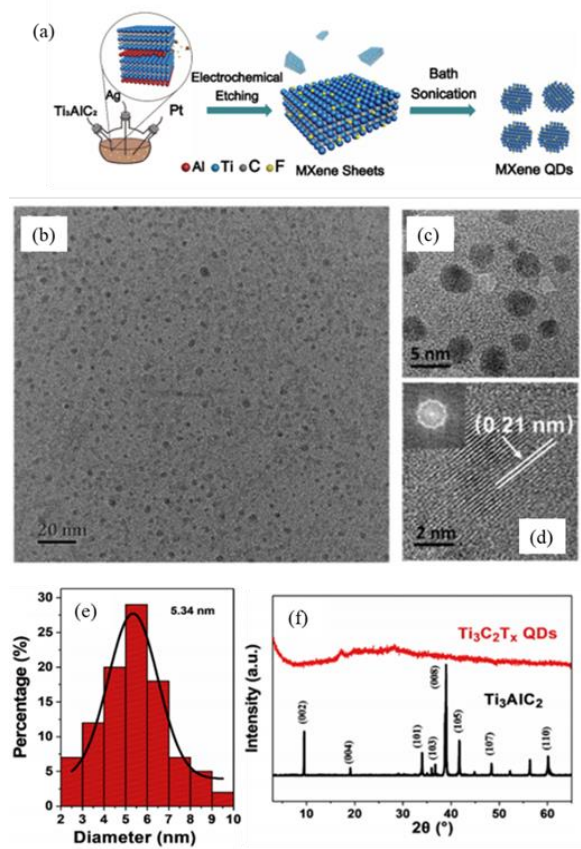

**Fig. 3:** (a) Synthesis setup, (b) TEM image, (c, d) HRTEM image, (e) diameter distribution, and (f) XRD pattern of $Ti_3C_2T_x$ QD, F. Yang et al. [64]. © American Chemical Society 2016.

A simple synthesis process of $Ti_3C_2T_x$ solution could also be done via stirring and ultrasonic vibration [102]. Firstly, a 40 mL of 0.008 g/mL PVA solution was stirred with 10 mg of $Ti_3AlC_2$ powder inside a beaker for 24 hours. Then, ultrasonication was performed to break the agglomerated $Ti_3AlC_2$ apart by cavitation for 2 hours. Finally, the supernatant was separated from the solid residue to complete the synthesis of $Ti_3C_2T_x$ solution. Aside from ultrasonic vibration, the $Ti_3C_2T_x$ nanosheet could also be synthesized through LPE method [103]. Firstly, 20 mg of $Ti_3C_2T_x$ powder was dispersed into 20 mL of ethanol in a water bath with temperature lower than 20 °C. This dispersion was then treated with ultrasonication using ultrasonic power of 200 W for 12 hours. Next, the $Ti_3C_2T_x$ suspension was centrifuged at 3000 rpm for 30 minutes to remove large bulks and the supernatant of $Ti_3C_2T_x$ nanosheet was finally obtained. Multilayer crystal structure $Ti_3C_2T_x$ with accordion-like morphology is observed via the SEM image in Fig. 4(a). Fig. 4(b) and 4(c) present the TEM images of the exfoliated $Ti_3C_2T_x$ nanoflakes with interlayer distance of 1 nm. The HRTEM image of the $Ti_3C_2T_x$ nanoflakes is depicted in Fig. 4(d), showing that the $Ti_3C_2T_x$ atoms are arranged and distributed in hexagonal shape with an equiangular lattice spacing of 0.26 nm.

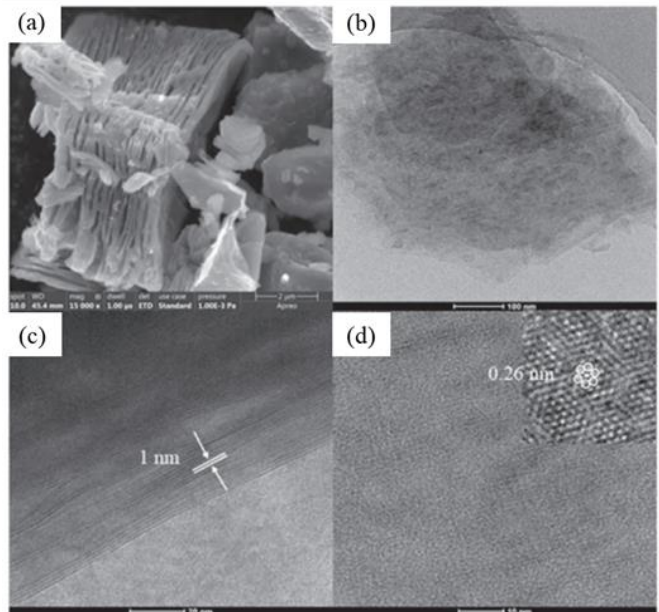

**Fig. 4:** (a) SEM image of multilayer $Ti_3C_2T_x$, (b) TEM image with 100 nm scale, (c) Interlayer distance of the $Ti_3C_2T_x$ nanoflakes measured as 1 nm, (d) HRTEM of the $Ti_3C_2T_x$ nanoflakes, X. X. Yan et al. [103]. © The Japan Society of Applied Physics 2021.

## 2.2 $Ti_3CNT_x$

The material properties of $Ti_3CNT_x$ are close to $Ti_3C_2T_x$ [104]. The schematic structure of monolayer $Ti_3CNT_x$ was presented with different surface terminations [51], as illustrated in Fig. 5(a) and 5(b). A mix arrangement of C and N in $Ti_3CNT_x$ was constructed by replacing one of four carbon atoms with nitrogen atom. This contributes to more dominant metallic characteristics of monolayer $Ti_3CNT_x$ in comparison to other semiconducting 2D materials based on the corresponding electronic band structures shown in Fig. 5(c) and 5(d). The stacking of monolayer $Ti_3CNT_x$ is primarily attributed to the hydrogen bond instead of Van der Waal's attraction. These hydrogen bonds are created between surface functional groups or indirect hydrogen bonds through intercalated water molecules [53, 105]. The electronic band structures of stacked $Ti_3CNT_x$ are preserved such as those observed in monolayer $Ti_3CNT_x$. This denotes the feasibility to well-functioning the SA without the needs of complex process to achieve monolayer solution.

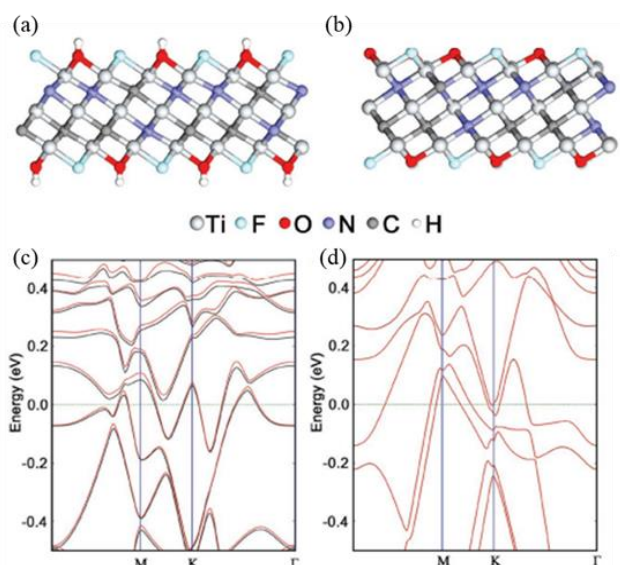

**Fig. 5:** The schematic structures of monolayer Ti3CNTx with surface terminations of (a) fluorine and hydrogen groups and (b) fluorine and oxygen groups, (c)-(d) the correspond electronic band structures, J. Li et al. [51]. © WILEY-VCH Verlag GmbH & Co. KGaA, Weinheim 2017.

The other $Ti_3CNT_x$ MXene was prepared by etching the Al layers from the raw material, $Ti_3AlC_2$ using 30% HF solution [51]. Firstly, 2 grams of $Ti_3CNT_x$ powder was mixed with 20 mL of 30% aqueous HF solution at room temperature which was then stirred with a magnetic bar for 18 hours. After that, the mixture was rinsed with DI water, stirred for 1 minute, and centrifuged at 3500 rpm for 3 minutes. This process was repeated for 5 cycles, followed by decanting until the supernatant possesses pH value of more than 6. Next, the $Ti_3CNT_x$ sediment was mixed with an aqueous solution of 54-56% tetrabutylammonium hydroxide (TBAOH) using a ratio of 100:1 to delaminate the multi-layered $Ti_3CNT_x$. The mixture was then stirred for 4 hours in room temperature. After that, the mixture was centrifuged at 3500 rpm for 2 minutes, followed by decanting the supernatant. Next, DI water was added with a very mild shake to redisperse the sediment, followed by centrifugation and decanting the supernatant for three times to eliminate the residue TBAOH. Then, 100 mL of DI water was added to the residue. The mixture was sonicated for an hour and centrifuged at 3500 rpm for another hour. Finally, the supernatant with mono- to few-layer $Ti_3CNT_x$ was synthesized. The optical absorption coefficient was calculated for both monolayer $Ti_3CNT_x$ and their stacked systems which cover a wide wavelength from 1000 to 3500 nm.

## 2.3 V₂CTₓ

$V_2CT_x$ is a $V_2C$-based MXene with featureless absorption spectrum covering visible to near-IR region with absorption coefficient of approximately $1.22 \times 10^5$ cm$^{-1}$ at 550 nm [106]. This contributes to twice transparency of $V_2CT_x$ than previous Ti-based MXene as a good conductor. The effective nonlinear absorption coefficient measured as -10$^{-21}$ m$^2$/V$^2$ also concludes the feasibility of $V_2CT_x$ as optical switches [107]. However, this MXene could face difficulties in synthesizing highly pure $V_2CT_x$ because the formation energy of $V_2C$ from raw material, $V_2AlC$ at 2.982 eV is considered moderate than other MXenes in $M_2X$ system [108]. Hence, it is cumbersome to attain delaminated flakes as $V_2CT_x$ does not contains strictly crystalline structure [109]. The $V_2CT_x$ was prepared through aqueous acid etching process by etching the parent $V_2AlC$ phase in HF and intercalated through TBAOH as shown in Fig. 6(a) [106]. The excess TBAOH was washed away and the mixture was sonicated in DI water to produce the colloidal suspension. Next, the suspension was spin casted to synthesize $V_2CT_x$ film with thickness of 400 nm. Fig. 6(b) shows the top-view of SEM image for a smooth spin-casted film. The bottom left inset illustrates the cross-section SEM image with film thickness of 400 nm, whilst the right inset shows the SEM image of the glass side spin-casted with various film thicknesses. The SEM image of the cross-section stacking structure of the $V_2CT_x$ made by filtration is portrayed in Fig. 6(c). The inset of this figure shows the flexibility and free-standing characteristics of the $V_2CT_x$ film. Fig. 6(d) shows the high magnified TEM image of the $V_2CT_x$ flakes obtained by drop casting colloidal suspensions produced using TBAOH. The flakes show high quality, soft and flexible with several overlapped and folded flakes as indicated by red arrows. The selected area electron diffraction (SAED) pattern of these flakes is depicted in the inset of Fig. 6(d). Based on the UV-VIS-NIR transmittance curve in Fig. 6(e), the $V_2CT_x$ shows broad absorption spectrum ranging from 500 to 2700 nm with free of distinct transition, especially for a thinner $V_2CT_x$ film such as 10 nm thickness dimension.

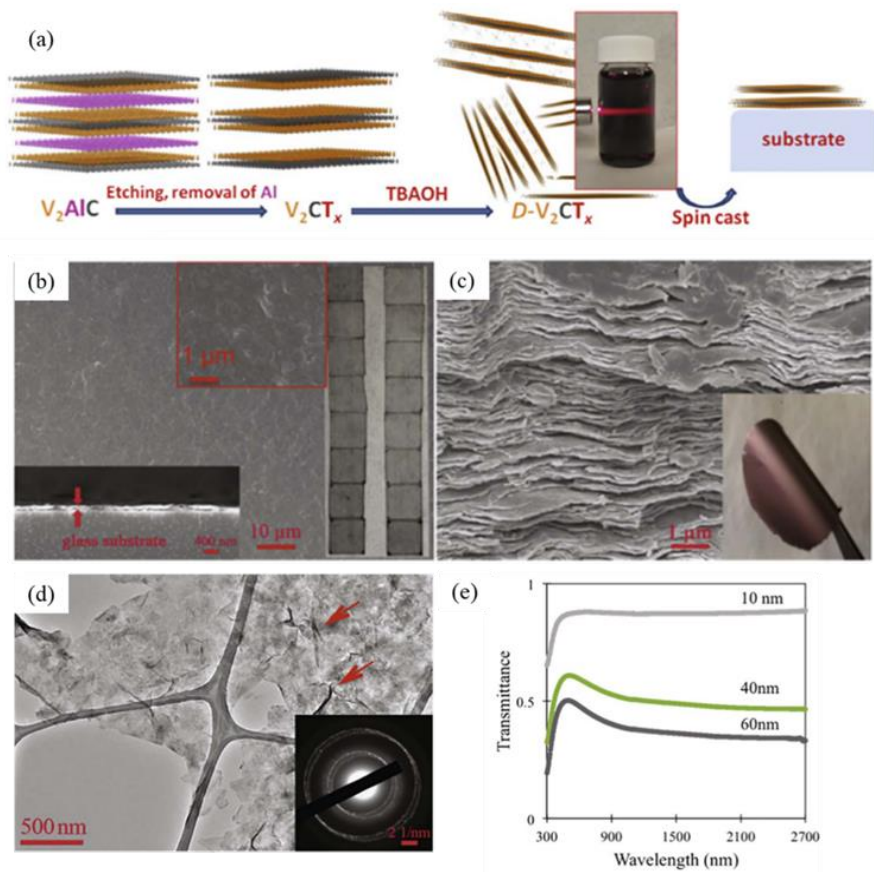

**Fig. 6:** (a) Schematic diagram to prepare $V_2CT_x$ followed to spin-cast the $V_2CT_x$ film, (b) top view and (c) cross-sectional view of SEM images for $V_2CT_x$ film, (d) TEM image of $V_2CT_x$ flake with the SAED pattern in the inset, (e) UV-VIS-NIR transmittance curve as a function of different $V_2CT_x$ film thickness, G. Ying et al. [106]. © Elsevier 2018.

Another example to prepare the $V_2CT_x$ powder was through typical aqueous acid etching method [40]. First and foremost, the $V_2C$ powder was stirred continuously in a mixture of 2.0 g of $V_2AlC$ (200 mesh) with 40 mL HF acid for 48 hours at 35 °C. Then, the mixture was diluted with DI water and centrifugation took place at 5000 rpm for 10 minutes per cycle for several times until the pH value of the supernatant is more than 6. The etched $V_2AlC$ was collected by filtration using PVDF membrane (0.45 mesh) and rinsed with 2 litres of DI water. Then, bath sonication was deployed to delaminate the $V_2AlC$ using a built-in water-cooling system at 400 W and constant temperature of 10 °C for 30 minutes. Finally, the dispersion was centrifuged at 5000 rpm for 30 minutes to produce the supernatant of $V_2CT_x$ nanosheets. Further centrifugation of this supernatant was conducted at 18000 rpm for another 30 minutes to obtain the precipitate. This precipitate was kept in a vacuum condition at 80 °C for 24 hours. Fig. 7(a) shows the SEM image of the MAX phase of $V_2CT_x$ which reveals the basal planes fan out and spread apart after the removal of Al from $V_2AlC$ with HF etchant. Based on the TEM image in Fig. 7(b), the exfoliated nanosheet exhibits a thin lateral range from ~50 to ~200 nm. The crystal structure was verified through SAED which exhibits a set of clear lattice fringes with inner plane spacing of 0.24 nm as depicted in Fig. 7(c). In addition, the thickness of the synthesized $V_2CT_x$ nanosheet was measured as ~11 nm, which is about 11 layers of $V_2C$ based on the AFM measurement in Fig. 7(d) [110, 111]. The XRD patterns of $V_2CT_x$ nanosheet in Fig. 7(e) denotes that the (002) peak at 8.6° of $V_2C$ appears after the Al-layer was etched by HF acid [112]. The absence of obvious XRD signal after exfoliation indicates the etched $V_2C$ has been sufficiently exfoliated. The synthesized $V_2CT_x$ nanosheet exhibits a broadband operation ranging from 280 to 2000 nm according to the UV-VIS-NIR spectrum as depicted in Fig. 7(f). Another work to synthesis $V_2CT_x$ nanosheet was proposed by selective etching of Al element from $V_2AlC$ [113]. The HF etching procedure is similar to [40], the only difference is the continuous stirring was conducted at 30 °C instead of 35 °C.

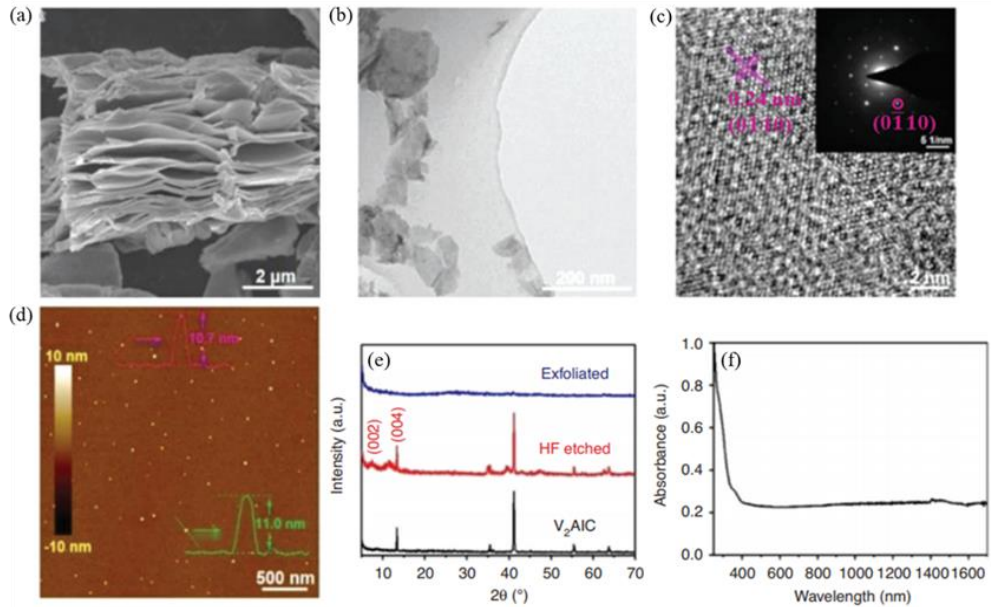

**Fig. 7:** (a) SEM image, (b) TEM image, (c) SAED image, (d) AFM measurement, (e) XRD patterns, and (f) UV-VIS-NIR spectrum of the synthesized $V_2CT_x$ nanosheet, C. Ma et al. [40]. © De Gruyter 2020.

## 2.4 Nb$_2$C

The research of Nb$_2$C could be traced back to 2013 when Nb$_2$AlC ceramic was synthesized by Naguib et al. [114]. According to the linear absorption spectrum measured with the Tauc method, the optical bandgap of Nb$_2$C is deduced as 0.81 eV [115]. Additionally, the NLO properties of the Nb$_2$C MXene were investigated, such as the imaginary part of third order susceptibility was studied as ~$10^{-11}$ to $10^{-10}$ e.s.u. and the nonlinear absorption coefficient was inferred as the values from -0.35 to 0.30 cm/GW when the wavelength is red-shifted from 1064 to 1600 nm [115]. The Nb$_2$C shows huge reduction of lattice thermal conductivity attribute to the approximate intensities between abnormal electron-phonon scattering and phonon-phonon scattering [116]. A high photothermal conversion efficiency and strong absorption were demonstrated for Nb$_2$C in near-IR region [117]. The Nb$_2$C could be synthesized through acid etching method by treating Nb$_2$AlC with 50% HF solution at room temperature [73]. After the etching process, the superfluous HF was rinsed repeatedly with DI water and centrifuged at 3000 rpm until the pH is greater than 6. Then, filtration was performed to collect the Nb$_2$C suspension through cellulose membrane. Next, the as-synthesized multilayer Nb$_2$C was dispersed in 30 mL of 25% tetrapropylammonium hydroxide (TPAOH) solution and stirred vigorously for 3 days at room temperature. Subsequently, the TPAOH was removed by washing the mixture with DI water and centrifuged at 15000 rpm for 10 minutes. The precipitate was then redispersed in dimethylformamide (DMF) and ethanol, and centrifuged at 3000 rpm for 10 minutes to separate Nb$_2$C colloidal from undelaminated Nb$_2$C particles. Finally, few-layered Nb$_2$C nanosheet was synthesized and 5 °C storage condition is required to extend its lifetime. Fig. 8(a) shows the SEM image of the Nb$_2$C nanosheet with accordion-like structure denoting the removal of Al layers from the MAX phase. The TEM image shown in Fig. 8(b) shows the 2D morphology of the exfoliated Nb$_2$C whereas the lighter profile indicates the ultra-thin characteristics of the Nb$_2$C nanosheet. Next, the crystal structure of the Nb$_2$C nanosheet was characterized with the hexagonal symmetry structure of the SAED pattern (inset) based on HRTEM image as depicted in Fig. 8(c). Subsequently, Fig. 8(d) illustrates the size distribution of few-layer Nb$_2$C nanosheet based on the TEM image, whilst the diameter of the nanosheet is measured from 30 to 270 nm with average size between ~112 ± 50 nm. The thickness of the Nb$_2$C nanosheet was characterized with the AFM measurement. The thickness was measured from 2.4 to 2.9 nm, corresponds to a composition of approximately 5 layers [73]. Fig. 8(f) presents the absorption measurement of the Nb$_2$C in ethanol for both before and after exfoliation. In contrast to the unexfoliated Nb$_2$C, the exfoliated Nb$_2$C exhibits stronger absorbance due to the modification of morphology and reduction in material size [118].

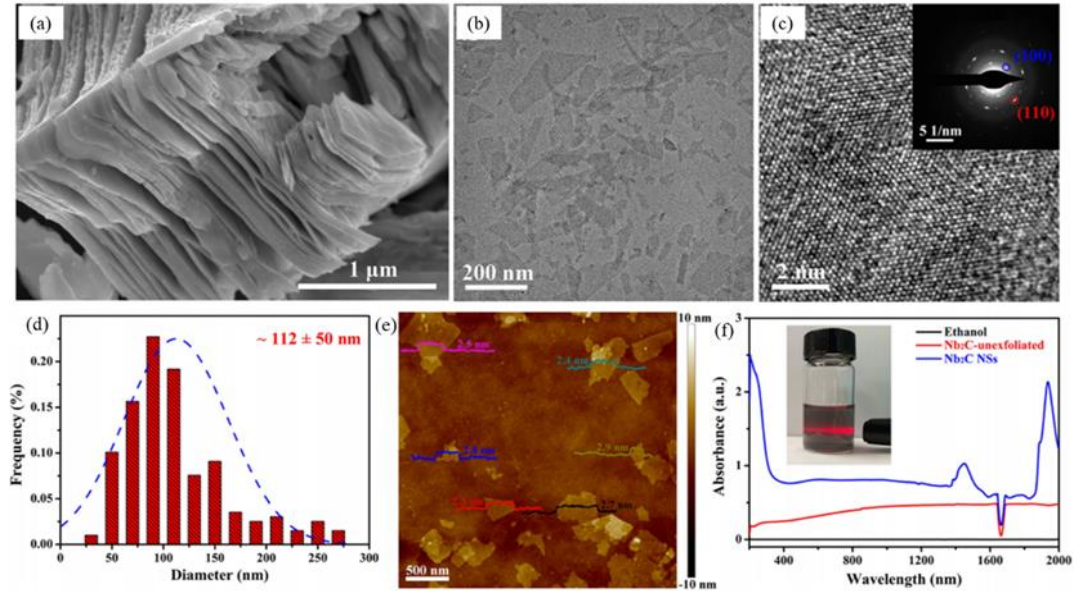

**Fig. 8:** (a) SEM image, (b) TEM image, (c) HRTEM image with SAED pattern, (d) size distribution, (e) AFM and (f) absorption measurement for as-synthesized Nb$_2$C nanosheet, L. Gao et al. [73]. © American Chemical Society 2021.

Besides that, the Nb$_2$C nanosheet could also be synthesized through LPE method [119, 120]. Firstly, 50 mg of Nb$_2$C powder was dispersed in 10 mL of IPA solution. Then, the mixture was sonicated at 40% power for four hours. Next, the undissolved flakes were removed through centrifugation of the suspension at 4000 rpm for 10 minutes, resulting in the homogeneous Nb$_2$C nanosheet in IPA solution. The as-synthesized Nb$_2$C nanosheet was characterized through the SEM image as shown in Fig. 9(a). The accordion-like structure denotes the whole grain was built up from multilayer Nb$_2$C nanosheet. In addition, the Raman spectrum of the Nb$_2$C nanosheet with three characteristic peaks is presented in Fig. 9(b). The 263 cm$^{-1}$ corresponds to the A$_{1g}$ mode attributed to the symmetrical out-of-plane vibration of Nb and C atoms [121, 122]. The other two peaks at 1350 cm$^{-1}$ and 1584 cm$^{-1}$ are contributed by the D and G bands of the carbon atoms [123, 124]. The other three peaks are associated with the Si substrate which were observed due to the penetration of scanning laser over the thin sample layer.

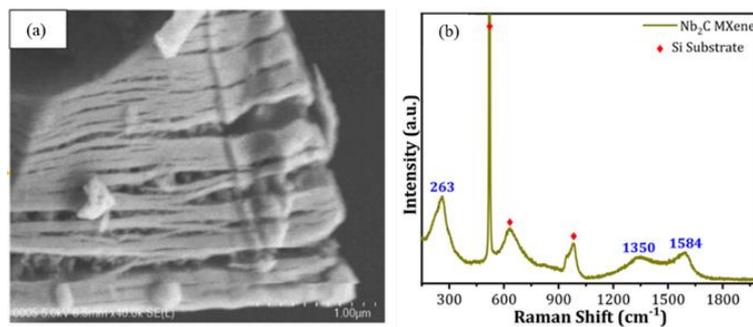

**Fig. 9:** (a) SEM image and (b) Raman spectrum of the as-synthesized Nb$_2$C nanosheet, H. Ahmad et al. [119]. © Elsevier 2021.

## 3 Saturable absorber

Flame-brushing technique is a popular method in tapered fiber fabrication [119]. A typical example for the flame-brushing technique is by melting the optical fiber with an oxy-LPG flame and pulling the optical fiber to reduce its waist diameter through a translational stage thus forming the tapered fiber as presented in Fig. 10(a) [125]. The tapered fiber induces evanescent field around its waist region for the interaction of absorbing materials with high heat dissipation mechanism [126, 127]. The SA could be formed on the tapered fiber platform in Figure 10(a) with several methods. Optical deposition is the most general technique in SA fabrication. For instance, a Hi-1060 fiber was tapered with a waist length of ~3 mm and waist diameter of ~5.2 μm [40]. A 980 nm pump was used for the optical deposition process and the V$_2$CT$_x$ nanosheets were dropped on the waist region of the tapered fiber whereas the adsorption of materials occurs due to the evanescent interaction. The whole process was observed through a power meter. The next tapering setup was done for SMF-28 under oxy-hydrogen flame and the tapering achieves ~13 µm waist diameter [63]. A similar optical deposition was deployed, with a light source of 60 mW and the volume of droplets are ~5 to 10 µL [63]. The optical deposition was stopped when the optical loss reaches 3 dB. Another tapered fiber with waist diameter of ~9.21 µm and insertion loss of 0.69 dB was employed for optical deposition for a $Ti_3C_2T_x$ suspension with 980 nm laser pump and the real-time power was controlled with power meter [64]. The deposition length is ~200 µm and insertion loss is ~1.39 dB. Moreover, a tapered fiber with waist diameter of 12 µm and waist length of 5 mm went through optical deposition with 980 nm pump laser for optical deposition of $Nb_2C$ SA via real-time observation [73]. The depth of the deposition took about 3 minutes. Some other examples include tapered fiber with waist diameter 15 µm for optical deposition of a $Ti_3C_2T_x$ SA [82], and also optical deposition of $Ti_3C_2T_x$ SA using a needle tubing to ensure the right amount of $Ti_3C_2T_x$ sample was injected on the tapered fiber [55].

Apart from optical deposition method, the SA was formed by spraying the $Nb_2C$ suspension on the waist region of a tapered fiber platform with waist diameter of ~3.56 µm on a hot plate at 60 °C, the solvent will evaporate once it hits the hot plate surface [119]. The optical loss of this SA is measured to be ~1.55 dB. Besides that, magnetron sputtering deposition (MSD) was employed to deposit few-layer $Nb_2CNNb_2C$ nanosheet on the tapered fiber [128]. The MSD method was referred from [129]. Firstly, concentrated HF and $H_2SO_4$ were used for hydrophilic treatment of fluorine mica (FM) to enhance the adhesion of film on the FM surface. Then, the FM sheet was rinsed repeatedly with alcohol and DI water. Next, the treated FM sheet was dried in an oven at 40 °C. Subsequently, the FM and MXene target were placed into a magnetron sputtering chamber system. Mechanical and molecular pumps were deployed to pump the chamber with vacuum degree of $6.8 \times 10^{-4}$ pa. Finally, the FM was rotated at a speed of 30 rpm to get a uniform MXene film. The MXene film was then sputtered on a microfiber substrate with waist diameter of ~18.03 µm with similar pressure and rotation speed.

D-shaped fiber is another type of microfiber which also induces evanescent interaction with the absorbing materials on its polished surface [130]. This side-polished fiber structure appears more robust than the tapered fiber [131]. Besides, the D-shaped fiber possesses numerous advantages such as long nonlinear interaction length, high optical damage threshold and strongly enhanced light-matter interaction [132, 133]. A detailed fabrication process of D-shaped fiber with mechanical technique was presented in [102]. The D-shaped fiber was fabricated with fiber diameter of 71.69 µm, core-cladding distance of 4.69 µm, and polishing length of 1400 µm, which was used for the drop-casting of $Ti_3C_2T_x$ solution to form a SA as presented in Fig. 10(b) [102]. Another D-shaped fiber with 6 µm polishing depth from the core was used for the dripping of $Ti_3C_2$ nanosheet dispersed with 0.1 mg NMP by ultrasonic bath [80]. Moreover, the stacked $Ti_3CNT_x$ monolayer film with thickness of 4-6 µm was attached to a D-shaped fiber with insertion loss and polarization dependent loss (PDL) of 4.5 dB and 1.8 dB, respectively at 1557 nm of wavelength [51]. The PDL is a measure of peak-to-peak optical power distribution by all possibilities of polarization states, which causes a variation in signal power and imbalance of optical signal-to-noise ratio of the transverse electric (TE) and transverse magnetic (TM) modes [134]. Subsequently, the TE and TM modes is the major effect of controlling the saturation level of the laser beam which causes variation in modulation depth of a SA [135]. Besides that, a D-shaped fiber with side-to-core depth of 6 µm was used for evanescent coupling of $Ti_3C_2T_x$ solution mixed with 6.67 mg/mL [13]. The insertion loss and PDL are 5.2 dB and 2.0 dB at 1900 nm, respectively. Inkjet printing implements the advantages of 2D materials for printed optoelectronic devices with small-footprint, ease of integration, geometry compatibility, scalability, and cost-efficient [81]. To fabricate a D-shaped SA, uniform and continuous $Ti_3C_2T_x$ nanosheet film was printed on the D-shaped fiber by setting the inter-droplet distance to be 35 µm. The film uniformity could be improved with multi passes due to the small inter-droplet distance.

Aside from polishing the D-shaped fiber using mechanical technique as aforementioned, an etched fiber could also be fabricated by dipping the optical fiber into 30% HF acid solution for 2 hours [136]. The HF etched away the silica glass part with removed coating. Then, the etched fiber was cleaned with pure water and ethanol to remove the remaining HF solution from the fiber surface. The etched fiber was then inserted into a protection sleeve, and $Ti_3C_2T_x$ solution was injected using a syringe. To seal the liquid $Ti_3C_2T_x$ inside the protection sleeve, both ends of the sleeve were melted using hot iron and sealed with glue gun. The end-product of this etched fiber SA is presented in Fig. 10(c). This SA structure is useful to improve the thermal damage threshold caused by physical contact and it is less sensitive to polarization attributed to the symmetrical geometry which is relatively more stable than D-shaped fiber [137]. Nonetheless, the utilization of etched fiber platform as the SA is less popular than its implementation in numerous types of sensors such as biosensor [138], refractometric sensor [139], and humidity sensor [140]. This is because the end-product of etched optical fiber is mostly nanometer-sized fiber tips which could not be easily achieved by either tapering or mechanical polishing approaches [141, 142]. In particular, the fabrication method of etched fiber is more haz-ardous than tapered fiber or D-shaped fiber due to the integration of HF solution which should not be handled continuously even wearing proper protection equipment. The other example of etched fiber was employed for the immersion into $Ti_3C_2$ solution to form the SA [143]. Through optical deposition with a laser light source, the $Ti_3C_2$ solution was attracted to the etched fiber surface. In contrast to the etched fiber, there is also a rather simpler method to form the SA by sandwiching the absorbing material such as $Ti_3C_2T_x$ between two fiber ferrules as depicted in Fig. 10(d) [103]. The $Ti_3C_2T_x$ was made into composite film by mixing 10 mL $Ti_3C_2T_x$ supernatant with 10 mL of PVA solution, which was treated by ultrasonication for 24 hours. The $Ti_3C_2T_x$/PVA composite film was cut into $1 \times 1$ mm$^2$ size and placed in between two fiber ferrules to form the SA. In comparison to microfibers, the fiber ferrule structure is relatively simpler for handling and operation at the expense of shorter interaction length and lower optical damage threshold due to the physical contact between the absorbing materials and the fiber ferrules [133, 144].

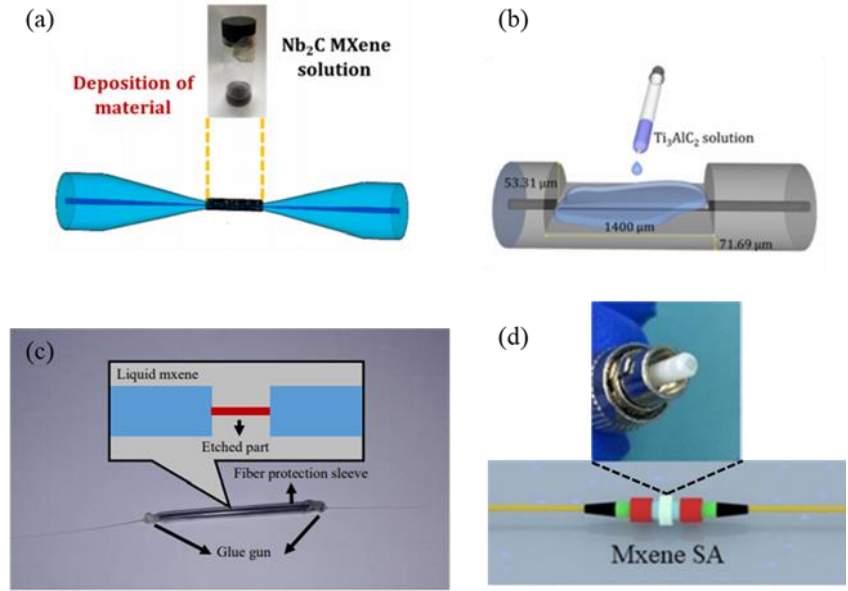

**Fig. 10:** SA with the structure of (a) tapered fiber, H. Ahmad et al. [119]. © Elsevier 2021. (b) D-shaped fiber, A. A. A. Jafry et al. [102]. © Optical Society of America 2020. (c) Etched fiber, J. J. Kim [136]. © Optical Society of Korea 2018. (d) Fiber ferrules, X. X. Yan et al. [103]. © The Japan Society of Applied Physics 2021.

A balance twin-detector measurement setup was proposed as a method to deduce the NLO response of the SA [63, 145]. The setup consists of an ultrafast laser source with the input power controlled by a variable optical attenuator, a coupler divides the optical signal into linear transmission without the SA and nonlinear transmission with the SA as shown in Fig. 11(a). The output powers of both linear and nonlinear routes were measured through optical power meters [40]. The nonlinear saturable absorption of a SA is calculated based on the following equation:

$$T(I) = 1 - [\alpha_s / (1 + I/I_{sat})] - \alpha_{ns}$$

where $T(I)$ is the transmission coefficient, $\alpha_s$ is the modulation depth, $I$ is the power of input light, $I_{sat}$ is the saturation intensity and $\alpha_{ns}$ is the non-saturable loss. The three important parameters, $\alpha_s$, $I_{sat}$, and $\alpha_{ns}$ could be measured from the nonlinear saturable absorption curve. An example of saturable absorption curve is presented in Fig. 11(b). In addition, the y-axis of this NLO curve could also be declared as transmission in a reverse manner depends on which parameters were employed as the sum of transmission and absorption is ~1 [146]. Table 1 summarizes the $\alpha_s$ and $I_{sat}$ according to different MXene SAs at different operating wavelength. Nevertheless, the $\alpha_{ns}$ is less likely mentioned in most of the reports, whereas only refs. [102] and [55] reported the values as 49.2% and ~73.52%, respectively. In the authors' opinion, this important parameter should not be neglected in the articles because low $\alpha_{ns}$ or $\alpha_s$:$\alpha_{ns}$ ratio are the indication of excellent SA quality [147, 148].

**Table 1:** Summary of nonlinear saturable absorption for MXene SAs at different wavelength.

| SA | Wavelength | $\alpha_s$ (%) | $I_{sat}$ | Ref. |
|---|---|---|---|---|
| $Ti_3C_2T_x$ | 1557.77 | 2 | 2.68 MW/cm$^2$ | [102] |
| $Ti_3C_2T_x$ | ~1550 | 0.98 | 256.7 MW/cm$^2$ | [55] |
| $Ti_3C_2T_x$ | 1558 | 11.3 | 1.94 mW | [63] |
| $Ti_3C_2T_x$ | ~1560 | 0.96 | 256.9 MW/cm$^2$ | [78] |
| $Ti_3C_2T_x$ | ~1550 | 3.6 | 2 MW/cm$^2$ | [103] |
| $Ti_3C_2T_x$ | 1) 1550 <br> 2) 2000 | 1) 29.3 <br> 2) 49 | 1) ~10.6 MW/cm$^2$ <br> 2) ~651.23 MW/cm$^2$ | [82] |
| $Ti_3C_2T_x$ | 1900 | ~15 | 18.6 MW/cm$^2$ | [13] |
| $Ti_3CNT_x$ | 1560 | 1.7 | 45W (Saturation power) | [51] |
| $Ti_3C_2$ | 1958 | 2.1 | ~20 mW | [80] |
| $Ti_2AlC$ | 1550 | 2.21 | 2.55 mW | [89] |
| $Nb_2C$ | 1560 | 12.94 | 0.0011 MW/cm$^2$ | [119] |
| $Nb_2C$ | ~1560 | ~2.5 | 0.92 GW/cm$^2$ | [128] |
| $Nb_2C$ | 1950 | ~6 | 0.4 MW/cm$^2$ | [120] |
| $V_2CT_x$ | 1064 | 23.7 | 1.46 mW | [40] |
| $V_2CT_x$ | 1558 | 48.8 | 0.5 mW | [113] |

An open aperture Z-scan system is another method to measure NLO properties of materials [149]. A typical Z-scan setup is illustrated in Fig. 11(c). Based on [52], three pulsed laser sources of 400 nm, 800 nm, and 1560 nm were deployed for the characterization of a $Ti_3C_2T_x$ SA. When the $Ti_3C_2T_x$ sample moves closer to the beam focus, the transmission gradually increased. This denotes the reduction in light absorption of the $Ti_3C_2T_x$ sample which gradually saturated with increased light intensity. The $\alpha_s$ and $I_{sat}$ were measured as 20% and 3.7 GW/cm$^2$ at 400 nm, 29% and 25 GW/cm$^2$ at 800 nm, and 41% and 7.3 MW/cm$^2$ at 1560 nm as shown in Fig. 11(d). Therefore, this measurement shows that $\alpha_s$ getting larger whilst $I_{sat}$ becomes smaller for $Ti_3C_2T_x$ at near-IR region. Moreover, another open aperture Z-scan measurement for $Ti_3C_2T_x$ SA with Ti:sapphire femtosecond laser oscillator and an optical parametric amplifier system with center wavelengths of 800, 1064 and 1550 nm was also conducted in [150]. The incident light beam was divided into 10% as the reference signal, whereas the remaining 90% beams are focused by a plano-convex lens that generates light and excites the sample. The resulting signal was finally collected by a power detector. The few layer $Ti_3C_2T_x$ nanosheet solution was placed in the cuvette and mounted on a horizontal axis translational stage. The $\alpha_s$, $I_{sat}$, and two photon absorption coefficients ($\beta$) were measured as 7.96 ± 0.29%, 224.62 ± 15.53 GW/cm$^2$, and – (2.32 ± 1.63) × 10$^{-2}$ cm/GW for 800 nm, 9.76 ± 0.27%, 215.97 ± 11.24 GW/cm$^2$, and – (2.41 ± 0.17) × 10$^{-2}$ cm/GW for 1064 nm, and 13.74 ± 0.80%, 258.88 ± 26.49 GW/cm$^2$, and – (0.66 ± 0.26) × 10$^{-2}$ cm/GW for 1550 nm at the highest pulse energy of 150, 160, and 180 nJ, respectively. Moreover, the open aperture Z-scan measurement was also demonstrated for $Ti_3C_2T_x$ SA with femtosecond lasers at center wavelengths of 540, 800, 1060 and 1550 nm [64]. The $I_{sat}$, $\alpha_s$, and $\alpha_{ns}$ was deduced as 1.26 ± 0.22 MW/cm$^2$, 2.94 ± 0.1%, and -7.32 cm/GW at 540 nm, 1.10 ± 0.04 MW/cm$^2$, 3.52 ± 0.38%, and -5.58 cm/GW at 800 nm, 0.216 ± 0.03 MW/cm$^2$, 7.72 ± 0.38%, and -4.80 cm/GW at 1060 nm, and 1.55 ± 0.257 MW/cm$^2$, 21.75 ± 1.14 %, and -22.04 cm/GW at 1550 nm, respectively. In contrast to twin-detector measurement setup which employs mostly optical fiber, the Z-scan method is more suitable for the integration of free-space optics due to the alignment of optical lenses to focus the laser beam on the material sample.

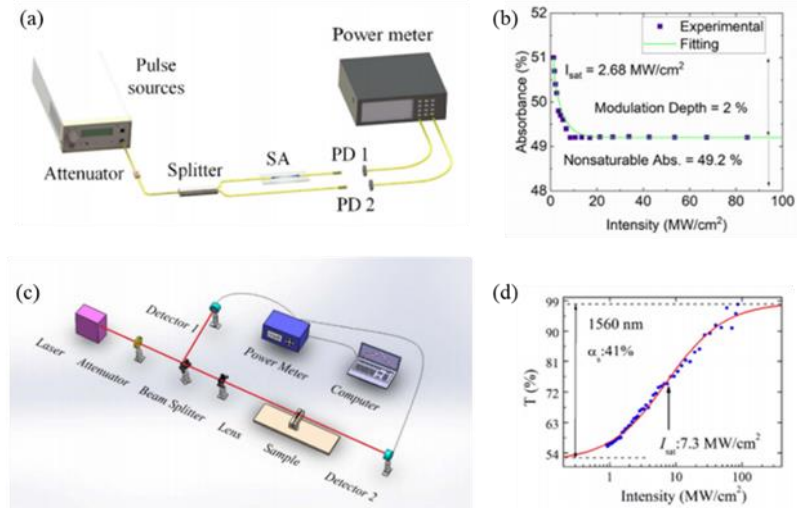

**Fig. 11:** (a) Balance twin-detector setup, Q. Wu et al. [63]. © Optical Society of America 2019. (b) nonlinear saturable absorption for MXene SA, A. A. A. Jafry et al. [102]. © Optical Society of America 2020. (c) Z-scan measurement setup and (d) nonlinear saturable transmission at 1560 nm for MXene sample, J. Li et al. [52]. © Chinese Laser Press 2019.

## 4 Near-IR mode-locked fiber laser

The implementation of SA into a laser cavity generates mode-locked fiber laser. For instance, a ring mode-locked ytterbium-doped fiber laser (YDFL) was demonstrated at a net GVD of normal dispersion, ~0.92 ps$^2$ [64]. The square or steep-sided spectral profile is a typical characteristic of this laser regime [151]. An example of the optical spectrum for the mode-locked YDFL is presented in Fig. 12(a). Apart from the YDFL, mode-locked erbium-doped fiber laser (EDFL) is the most popular laser regime. The ring-cavity mode-locked EDFL with net GVD of -0.221 ps$^2$ [119], -0.65 ps$^2$ [136], -0.7 ps$^2$ [143], and -2.5 ps$^2$ [102] were presented for soliton mode-locking. The Kelly's sidebands shown in Fig. 12(b) denotes the net anomalous GVD of the mode-locked EDFL, which is a typical characteristic of soliton operation of lasers [152]. In addition, a ring cavity mode-locked EDFL with net GVD of -0.14 ps$^2$ and 0.008 ps$^2$ were demonstrated for soliton and stretched pulses, respectively [63]. The pulse duration was measured as 597 fs for soliton mode-locking regime, and it is stretched to 104 fs by managing the GVD of the laser cavity. This stretched-pulse result was preliminary reported in [153]. The net GVD of the stretched pulse is close to zero which is very useful to achieve extremely short pulses such as 77 fs [154] and 88 fs [155]. Next, a ring cavity composed of mode-locked EDFL and thulium-doped fiber (TDFL) with net GVD of -0.66 ps$^2$ and -2.32 ps$^2$ were presented for $Ti_3C_2T_x$ SA [82]. Subsequently, TDFL was typically employed for ~2 μm emission. A ring cavity mode-locked TDFL with net anomalous dispersion [13], such as -1.046 ps$^2$ [156] was presented with $Ti_3C_2T_x$ SA and its optical spectrum is portrayed in Fig. 12(c). Moreover, 2 μm mode-locked fiber laser was also demonstrated with holmium-thulium-doped gain fiber and $Nb_2C$ SA [120].

Apart from fundamental pulses, harmonic mode-locking was also extensively demonstrated. The passively harmonic mode-locking is a technique to achieve higher repetition rate by $n^{th}$ order of the fundamental mode-locked laser pulse such that multiple pulses circulate in the laser cavity with equal spacing [157]. For instance, a ring cavity mode-locked EDFL with net anomalous GVD was presented to achieve 1.01 GHz at its $206^{th}$ order harmonic mode-locking [113]. The pulse train of this $206^{th}$ order harmonic mode-locking is deduced with repetition time of 1.001 ns as shown in Fig. 12(d). In [103], $56^{th}$ harmonic mode-locked EDFL was generated with pulse repetition rate of 622.2 MHz. Then, a ring cavity mode-locked EDFL with net anomalous GVD was presented to achieve 218.4 MHz at its $36^{th}$ order harmonic mode-locking [78]. The other ring cavity mode-locked EDFL with net GVD of -0.373 ps$^2$ was presented for soliton mode-locking using $Nb_2C$ SA, which was also demonstrated with a harmonic mode-locking at $69^{th}$ order for mode-locked TDFL [73]. Besides that, a ring cavity mode-locked TDFL with net GVD of -1.048 ps$^2$ were presented for either vector soliton (VS) at $19^{th}$ harmonic order or noise-like pulses (NLP) at fundamental order by adjusting the polarization controller [80].

Next, transition between self-started mode-locking and bound-soliton mode-locking was presented using $Ti_3C_2T_x$ SA by changing the pump power and adjusting the polarization controller [55]. According to Fig. 12(e), the presence of strong pedestal peaks aside the fundamental peak are the characteristics for bound-state mode-locking. The bound-state mode-locking or soliton molecules are observed due to the co-existence of several solitons in the laser cavity, whilst its build-up and transition dynamics were observed via the real-time measurement [158, 159]. Moreover, square pulse emission was presented which achieves pulse energy as high as 0.89 nJ [128]. In Fig. 12(f), the pulse width increased along with pump power at a constant repetition rate. This is a typical characteristic for square pulse emitted with dissipative soliton resonance regime [160]. An advantage of square pulse emission is its wave-breaking free output which maintains high pulse energy [161]. Moreover, the operation of mode-locked fiber laser was also reported at different operating wavelengths. For instance, dual-wavelength mode-locking was observed at center wavelengths of 1575.6 and 1587.5 nm with 3-dB spectral bandwidth of 3.1 and 3.4 nm, respectively as illustrated in Fig. 12(g) [103]. In addition, mode-locked fiber laser was also presented at far wavelength bands. For example, the $Ti_3C_2T_x$ SA was used for the ultrafast laser in 1064 nm and 1550 nm with net GVD of 0.35 $ps^2$ and -0.42 $ps^2$ [77]. Inkjet-printed $Ti_3C_2T_x$ SA was implemented for the ultrafast laser in 1064 nm and 1550 nm with net GVD of 0.227 $ps^2$ and -0.22 $ps^2$ [81]. Moreover, two laser cavities consist of a ring cavity mode-locked YDFL and EDFL with net normal and anomalous dispersion, respectively was presented in [150]. The multiple-band mode-locked lasers are useful for the practical applications such as high-precision dual-comb [162], dual comb spectroscopy and terahertz wave generation [163].

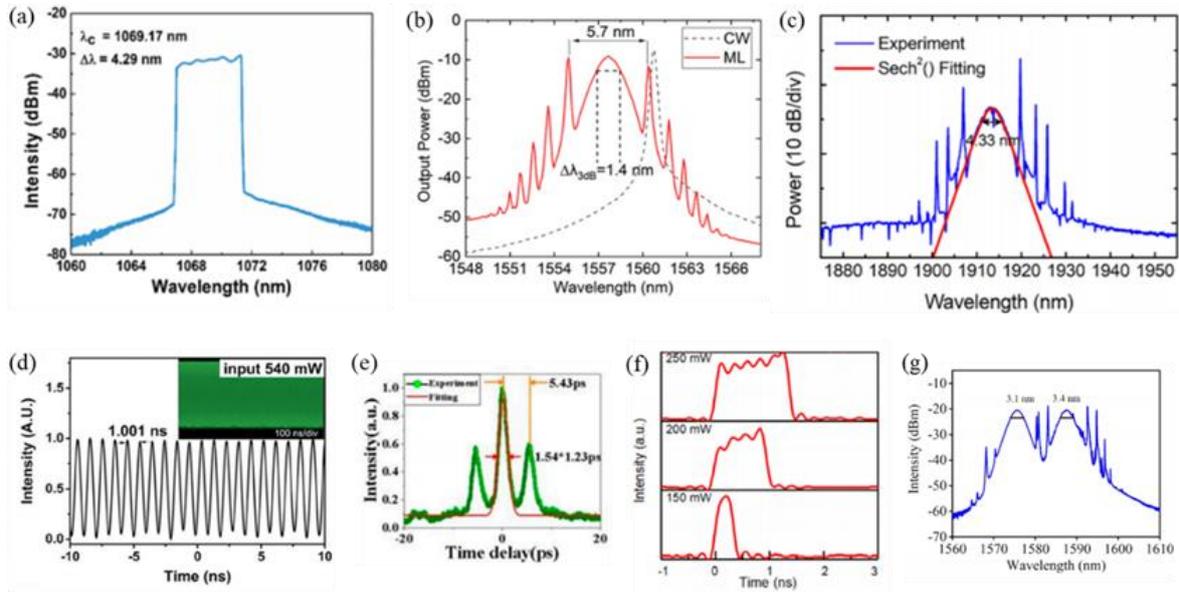

**Fig. 12:** Optical spectrum for typical mode-locked (a) YDFL, F. Yang et al. [64]. © American Chemical Society 2020. (b) EDFL, A. A. A. Jafry et al. [102]. © Optical Society of America 2020. (c) TDFL, Y. I. Jhon et al. [13]. © De Gruyter 2021. (d) Pulse train for harmonic mode-locking at 206[th] order, W. Huang et al. [113]. © De Gruyter 2020. (e) Autocorrelation trace for bound-state mode-locking with strong pedestal, C. Wang et al. [55]. © Wiley-VCH GmbH 2020. (f) Pulse train for rectangular pulse emission with broadened pulse width at constant repetition rate, G. Li et al. [128]. © Optical Society of American 2020. (g) Dual-wavelength emission for mode-locked EDFL, X. X. Yan et al. [103]. © The Japan Society of Applied Physics 2021.

Apart from the material-based SA, a ring-cavity hybrid mode-locked YDFL was demonstrated with fast SA (NPE) and slow SA ($V_2CT_x$ SA) that achieves several advantages such as low threshold of mode-locked laser, SNR of 13 dB higher than only NPE and pulse duration of 72 fs (GVD = ~0.1 $ps^2$) [40]. The function of slow SA is to self-start and stabilizes the mode-locked laser, whereas the fast SA shortens the pulse duration [164, 165]. Table 2 summarizes the technical result in the recent five years by integrating the SA into the mode-locked fiber laser in near-IR region with different fabrication techniques, materials, SA structure, deposition methods, and net cavity GVD. The $\lambda_c$, $\Delta\lambda$, $\Delta\tau$, TBP and SNR are the abbreviation of center wavelength, 3-dB spectral bandwidth, pulse duration, time bandwidth product, and signal-to-noise ratio, respectively.

**Table 2:** Summary of technical result for the mode-locked fiber laser.

| Year | Fabrication Method | Material | SA Structure | Deposition Technique | Net cavity GVD (ps$^2$) | $\lambda_c$ (nm) | $\Delta\lambda$ (nm) | $\Delta\tau$ | TBP | SNR (dB) | Repetition rate (MHz) | Ref. |
|---|---|---|---|---|---|---|---|---|---|---|---|---|
| 2017 | Aqueous acid etching | Ti$_3$CNT$_x$ | D-shaped fiber | Film attachment | Anomalous | 1557 | 5 | 660 fs | 0.4 | 60 | 15.4 | [51] |
| 2018 | - | Ti$_3$C$_2$T$_x$ | Etched fiber | Immersion | -0.65 | 1555.84 | 3.75 | 800 fs | - | 65 | 6.22 | [136] |
| 2018 | Aqueous acid etching | Ti$_3$C$_2$T$_x$ | D-shaped fiber | Solution deposition | 1. 0.35<br>2. -0.42 | 1065.89<br>1555.01 | 4.4<br>22.2 | 480 ps<br>159 fs | 557.7<br>0.45 | 56<br>62 | 18.96<br>7.28 | [77] |
| 2019 | Aqueous acid etching | Ti$_3$C$_2$T$_x$ | Tapered fiber | Optical deposition | -1.046 | 1862 | 1.86 | 2.11 ps | - | 65 | 13.45 | [156] |
| 2019 | - | Ti$_3$C$_2$ | Etched fiber | Immersion & optical deposition | -0.7 | - | ~1545 | 530 fs | - | - | 5.37 | [143] |
| 2019 | Aqueous acid etching | Ti$_3$C$_2$T$_x$ | D-shaped fiber | Spin coating | Anomalous | 1567.3 | 3.1 | 946 fs | 0.358 | 70.7 | 8.24 | [52] |
| 2019 | Aqueous acid etching | Ti$_3$C$_2$T$_x$ | Tapered fiber | Optical deposition | 1. ~-0.14<br>2. ~0.008 | 1564.24<br>1550 | 5.21<br>42.54 | 597.8 fs<br>104 fs | ~0.382<br>0.55 | 55.2<br>62.4 | 17.9<br>20.03 | [63] |
| 2019 | Aqueous acid etching | Ti$_3$C$_2$T$_x$ | D-shaped fiber | Inkjet printing | 1. 0.227<br>2. -0.22 | 1060<br>1550 | -<br>25.6 | 215 ps<br>114 fs | -<br>- | 63<br>65 | 16.07<br>11.76 | [81] |
| 2020 | Aqueous acid etching | Ti$_3$C$_2$T$_x$ | Tapered fiber | Optical deposition | 1. -0.66<br>2. -2.32 | 1564.44<br>1891.82 | 1.32<br>1.94 | 2.03 ps<br>2.18 ps | 0.328<br>0.354 | -<br>62 | 5.85<br>5.97 | [82] |
| 2020 | Aqueous acid etching + LPE | Ti$_3$C$_2$T$_x$ | Tapered fiber | Optical deposition | Anomalous | 1. 1566.9 (1$^{st}$)<br>2. 1566.9 (36$^{th}$) | 2.71<br>3.51 | 1.28 ps<br>850 fs | 0.424<br>- | 71<br>36 | 6.032<br>218.4 | [78] |
| 2020 | Electrochemical exfoliation | Ti$_3$C$_2$T$_x$ | Tapered fiber | Optical deposition | Normal | 1069.17 | 4.29 | 357 ps | 401.93 | 68 | 4.77 | [64] |
| 2020 | Aqueous acid etching | Ti$_3$C$_2$T$_x$ | Tapered fiber | Optical deposition | Anomalous | 1. 1566.1<br>2. 1566.1 (Soliton molecules) | 1.21<br>1.26 | 993 fs<br>1.23 ps | -<br>- | 49<br>60 | 6.087<br>6.087 | [55] |
| 2020 | Aqueous acid etching | V$_2$CT$_x$ | Tapered fiber | Optical deposition | -0.6 | 1. 1559.12<br>2. 1559.12 | 0.9<br>3.1 | 3.21 ps<br>940 fs | 0.351<br>0.359 | 57<br>55 | 4.7<br>1010 (206$^{th}$) | [113] |
| 2020 | Aqueous acid etching | V$_2$CT$_x$ | Tapered fiber | Optical deposition | 0.1 | ~1064 | 3 | 72 fs | - | 71 | 38.5 | [40] |

| Year | Fabrication Method | Material | SA Structure | Deposition Technique | Net cavity GVD ($ps^2$) | $\lambda_c$ (nm) | $\Delta\lambda$ (nm) | $\Delta\tau$ | TBP | SNR (dB) | Repetition rate (MHz) | Ref. |
|---|---|---|---|---|---|---|---|---|---|---|---|---|
| 2020 | - | $Nb_2CNNb_2C$ | Tapered fiber | MSD | -0.249 | 1567.5 | 3.2 | 0.33 - 2.061 ns (Square pulse) | - | 55 | 15.2 | [128] |
| 2020 | Stirring and ultrasonic vibration | $Ti_3C_2T_x$ | D-shaped fiber | Drop casting | -2.5 | 1557.63 | 1.4 | 2.21 ps | 0.363 | 67 | 1.89 | [102] |
| 2021 | LPE | $Nb_2C$ | Tapered fiber | Spraying and evaporation | -0.22 | 1559 | 4 | 770 fs | 0.38 | 66.5 | 14.12 | [119] |
| 2021 | Aqueous acid etching | $Nb_2C$ | Tapered fiber | Optical deposition | 1. -0.373  2. Anomalous | 1559.98  1882.9 (69th) | 4.6  2.16 | 603 fs  2.27 ps | 0.341  0.442 | 45  60 | 12.54  411 | [73] |
| 2021 | LPE | $Nb_2C$ | Tapered fiber | Spin coating | 1. -1.28  2. -1.17 | 1944  1950.8 | 2.8  3 | 1.67 ps  1.34 ns | 0.32  0.317 | 52  26.4 | 1. 9.35  2. 11.76 | [120] |
| 2021 | LPE | $Ti_3C_2T_x$ | Fiber ferrule | Film attachment | -0.38 | 1. 1573.2 (1st)  2. 1590.7 (56th)  3. a) 1575.6  3. b) 1587.5 | 2.7  1590.7  a) 3.1  b) 3.4 | 1.73 ps  1.31 ps  -  - | 0.566  0.459  -  - | 62.2  33.4  -  - | 11.1  622.2  -  - | [103] |
| 2021 | Aqueous acid etching | $Ti_3C_2T_x$ | D-shaped fiber | Solution deposition | 1. Normal  2. Anomalous | 1037.8  1530.85 | -  11.1 | 792 ps  265 fs | -  0.377 | 75  47 | 16.5  8.46 | [150] |
| 2021 | Aqueous acid etching | $Ti_3C_2T_x$ | D-shaped fiber | Drop casting | Anomalous | 1913.7 | 4.33 | 897 | 0.318 | 66 | 16.77 | [13] |
| 2021 | Aqueous acid etching | $Ti_3C_2$ | D-shaped fiber | Dripping | -1.048 | 1. 1964 (NLP)  2. 1965 (VS) | 3.3  4 | -  - | -  - | 56  69 | 13.7 (1st)  260.7 (19th) | [80] |
| 2021 | - | $Ti_2AlC$ | Fiber ferrule | Film attachment | Anomalous | 1559 | - | 680 | - | 64.6 | 5.16 | [89] |

# 5  Challenges and recommendations

Metallic conductors are generally poor SA [13]. The introduction of MXene with excellent nonlinear saturable absorption and flexible combination of $M_{n+1}X_nT_x$ contribute to the realization of excellent SA with various possibilities. Therefore, numerous unexploited MXenes with well-studied material properties are the future trend to generate mode-locked fiber laser. Based on the previous studies, the bandgap of the MXene was primarily predicted such as less than 0.2 eV for $Ti_3C_2T_x$ [43] and 0.81 eV for $Nb_2C$ using the Tauc method [115]. The significance of bandgap is shown in matching the operating wavelength of the mode-locked laser. Hence, a more relevant experimental bandgap measurement for the MXenes is crucial to validate the theoretical prediction. Apart from bandgap measurement, the synthesis of MXene also shows certain limitation. For instance, aqueous acid etching process is frequently employed for the synthesis of MXene. HF solution is very acidic which requires careful handling procedures and additional protection during the synthesis process. The improper management of residue HF solution will also result in severe pollution. Instead of using aqueous acid etching process, electrochemical exfoliation is a more viable and safer option that implement highly fluorinated, non-aqueous ionic liquid as the electrolyte to prepare MXene with enhanced stability [64]. In addition, the storage of MXene requires low temperature condition, such as below 5 °C to extend the lifetime [73]. Therefore, an obvious problem of MXene is its environmental stability and durability to preserve its initial performance. This problem is worthy for further investigation to extend its lifetime and reliability, especially by storing the MXene in room temperature.

Based on Table 2, tapered fiber is among the most deployed fiber platform to fabricate the SA. Nonetheless, most articles reported on the dimension of the tapered fiber without deeply studying the influences of adiabaticity of the tapered fiber. For instance, an optimization between the strong evanescent field and low loss tapered fiber with minimal length could be an interesting research study [166, 167]. In terms of the nonlinear saturable absorption properties of the MXene, most MXene SAs did not discussed on the non-saturable loss. This parameter is important to describe the properties of the MXene SA. For instance, mode-locked laser could not be achieved with a high non-saturable loss SA even if the modulation depth is sufficient. An example was shown in the transmission of $V_2CT_x$ SA that reduced from 89% to 27% when the film thickness increases from 11 nm to 116 nm [106]. Subsequently, the non-saturable loss of 73% in 116 nm thick $V_2CT_x$ might not be sufficient to generate an efficient fiber laser. Therefore, the description of non-saturable loss for the MXene SA should also be emphasized in the future article. The design of GVD to achieve conventional soliton, stretched pulse, self-similar soliton, and dissipative soliton is a well-known technique. There are also some other popular regimes, such as harmonic mode-locking, soliton molecule, and square pulse emission. The result in mode-locked fiber laser to achieve shorter pulse duration in sub-tens of femtosecond is much challenge unless the chirped pulse is compressed in dissipative soliton regime. In addition, the research would be more interesting if the physics for the formation of mode-locked laser from raised relaxation state, Q-switching and other transient dynamics are studied with dispersive Fourier transform [168], time lens [169], and frequency resolved optical gating techniques [170].

# 6  Conclusion and outlook

In short, this review outlines the recent development of MXene SA to achieve mode-locked fiber laser in near IR region. Although MXene is relatively newer than CNT, graphene, TMD and other materials as the SA, its excellent nonlinear saturable absorption properties are very promising to behave as a viable SA candidate. Up to now, a few MXene materials have been proposed as the SA, whereas the most popular MXene SA is $Ti_3C_2T_x$. This is due to its narrow bandgap, high nonlinear absorption coefficient, astounding modulation depth as high as ~50%, and larger effective nonlinear absorption coefficient than other 2D materials. The synthesis and material characterization of $Ti_3C_2T_x$ and other MXenes were then discussed. Based on the review, aqueous acid etching process is most frequently employed for the synthesis of MXene. Next, these MXene was integrated on the fiber platform as the SA. Microfiber was demonstrated in the structures of tapered fiber, D-shaped fiber or etched fiber which exhibits excellent heat dissipation mechanism and high optical damage threshold. These SA materials were then examined with the nonlinear saturable absorption characteristics using either twin-detector or Z-scan measurement systems. From these measurements, the modulation depth and saturation intensity of each SAs were deduced. Finally, these SAs were implemented into the fiber laser cavity to generate mode-locking in near IR region. A summary of the overall research articles was tabulated in Table 3. Subsequently, the issues and challenges met in this research area were discussed and several recommenda-tions were provided to address these problems thus rendering its potential for future development of this research topic. Finally, this review is expected to provide a better insight to the readers on the recent development of MXene SA in mode-locked fiber laser and its future perspectives in both scientific researches and industrial applications.

# Acknowledgement

This work was funded by the National Natural Science Foundation of China (62035010 and 11774310) and the Natural Science Foundation for Distinguished Young Scholars (61525505).

# Author contributions

All authors have accepted responsibilities for the entire content of this manuscript and approved its submission.

**22**